# Calculation of dispersion equations for uniaxial dielectric-magnetic mediums

Omid Khakpour[a], Seyed Mojtaba Rezaei Sani[b]

[a]Department of Physics, Payame Noor University (PNU), P. O. Box 19395-3697 Tehran, Iran

[b]School of Nano Science, Institute for Research in Fundamental Sciences, P. O. Box 19395-5531, Tehran, Iran

**ABSTRACT**

In this overview paper, we investigate the dispersion of electromagnetic waves for dissipative and non-dissipative dielectric-magnetic uniaxial mediums. Changing the sign of one component of dielectric permittivity ($\varepsilon$) or magnetic permeability ($\mu$) from positive to negative will lead to sixteen different cases (real or imaginary) for a non-dissipative medium and eight different cases for dissipative one. In a non-dissipative medium, dispersion relations follow the elliptic/hyperbolic relations. This rule completely vanishes regarding the dissipative medium. While the number of dispersion cases in a dissipative medium is lower than a non-dissipative one, there are cases which are not allowed in non-dissipative mediums but allowed in dissipative mediums.

## 1. INTRODUCTION

In the past decade, new structured electromagnetic materials which in some frequency ranges have both negative dielectric permittivity ($\varepsilon$) and magnetic permeability ($\mu$) have been developed [1-5]. Negative $\varepsilon$ and $\mu$ gives rise to negative refractive index materials which can exhibit exotic and unique electromagnetic properties not inherent in the individual constituent components [6]. The concept was first introduced by Veselago in 1968 [7] which theoretically investigated the electrodynamic consequences of a medium having both $\varepsilon$ and $\mu$ negative and concluded that such a medium would have dramatically different propagation characteristics stemming from the sign change of the group velocity, including reversal of both the Doppler shift and Cherenkov radiation, anomalous refraction, and even reversal of radiation pressure to radiation tension.

The change in the sign of refractive index has been predicted to lead to a variety of unique electromagnetic phenomena [8] one of the best examples is the perfect lens that enables imaging with sub-wavelength image resolution [9-11]. Artificial structures with effectively negative real permittivity and permeability -called metamaterials- are developing by scientists working in optics, electromagnetism, physics, engineering, and materials science. For example, Pendry *et al.* [1] showed that microstructures built from nonmagnetic conducting sheets exhibit an effective magnetic permeability $\mu_{eff}$, which can be tuned to values including large imaginary components of $\mu_{eff}$. Also Smith *et al.* [2] demonstrated that a periodic array of interspaced conducting nonmagnetic split ring resonators and continuous wires can exhibits a frequency region in the microwave regime with simultaneously negative values of effective permeability and permittivity.

The dispersion relations for conventional uniaxial dielectric-magnetic mediums may be characterized as elliptical or elliptical–like, according to whether the medium is non-dissipative or dissipative, respectively. In this paper, in order to investigate the behavior of negative refractive index materials we study the issues of electromagnetic waves propagation in metamaterials and effect of dissipation in the propagating medium with a particular focus on the uniaxial medium, where its distinguishing feature is its axis of symmetry called optic axis. We will show that the dispersion relations in a non-dissipative medium turn out to be in the form ellipsoids of revolution, or have representations of hyperboloid of one sheet or hyperboloid of two sheets; but, in the case of a dissipative medium this order vanishes. In general, according to the signs of elements of permittivity and permeability tensors, sixteen possible states are obtained for propagation in a non-dissipative electromagnetic medium. In dissipative mediums, the number of states is reduced and eight ones remain. There are forbidden conditions where there are no possible relations for non-dissipative imaginary parts, but under the same condition, there is an answer for the dissipative state.

## 2. THEORETICAL ANALYSIS

Propagation of electromagnetic plane waves in materials at a fixed frequencies of the wave vector causes the dispersion surfaces. For our theoretical analysis of wave propagation we choose an optic axis that is in the direction of the unit vector $\hat{\imath}$. Then, we calculate the dispersion relations and discuss the issue for both cases of dissipative and non-dissipative mediums. To simplify the analysis we assume a linear material with permittivity $\underline{\underline{\varepsilon}}$ and permeability $\underline{\underline{\mu}}$ tensors which are simultaneously diagonalizable [12, 13]. i.e.

$$\underline{\underline{\varepsilon}} = \begin{bmatrix} \varepsilon_\parallel & 0 & 0 \\ 0 & \varepsilon_\perp & 0 \\ 0 & 0 & \varepsilon_\perp \end{bmatrix} \quad and \quad \underline{\underline{\mu}} = \begin{bmatrix} \mu_\parallel & 0 & 0 \\ 0 & \mu_\perp & 0 \\ 0 & 0 & \mu_\perp \end{bmatrix} \qquad 1$$

where $\varepsilon_\parallel$ and $\mu_\parallel$ are the elements of permittivity and permeability tensors along the optic axis; and $\varepsilon_\perp$ and $\mu_\perp$ are the elements of the two tensors in the plane perpendicular to the optic axis, respectively. For non-dissipative mediums elements of these two tensor are real whereas for dissipative ones they are complex-valued. The real parts of these scalars are positive for natural mediums, but they can have any sign for metamaterials. There is no fundamental objection to the real parts of the $\varepsilon$ and $\mu$ being negative. The dispersion equation for plane waves in such mediums leads to the propagation of two different types of linearly polarized waves, called magnetic and electric modes [14, 15].
The electric and magnetic waves are

$$\begin{aligned} E(r) &= E_0 \; e^{iK.r} \\ H(r) &= H_0 e^{iK.r} \end{aligned} \qquad 2$$

where $K$ wave vector is defined in the most general form

$$K = \alpha \hat{\imath} + \beta \hat{k} \qquad 3$$

where α ∈ R, β ∈ C and $\hat{k}$ is the unit vector directed along the z axis [16]. This form of k is appropriate to planar boundary value problems [16] and from the practical viewpoint of potential optical devices [17]. We note that the plane waves (2) are generally non-uniform.

So the form of the electric field will be as follows

$$E(r) = E_0 \; e^{i\alpha x} \; e^{iRe\{\beta\}z} e^{-Im\{\beta\}z} \qquad 4$$

To calculate the dispersion surfaces, we seek for a relation between $\alpha$ and $\beta$. Plane wave propagation in a uniaxial medium is characterized in terms of a dispersion relation which is quadratic in terms of the corresponding wave vector components. By solving the eigenvalue equation, the dispersion relation for the extraordinary waves is obtained. The goal is to find a more complete calculation of the dispersion surfaces in a dielectric-magnetic dissipative or non-dissipative medium. Here we analyze the dispersion surfaces for a dielectric-magnetic medium. The source–free Maxwell curl postulates

$$\nabla \times \boldsymbol{E}(\mathbf{r}) = i\omega \underline{\underline{\mu}} \boldsymbol{H}(\mathbf{r}) \tag{5}$$

$$\nabla \times \boldsymbol{H}(\mathbf{r}) = -i\omega \underline{\underline{\varepsilon}} \, \boldsymbol{E}(\mathbf{r}) \tag{6}$$

Combining (5) and (6) we have

$$\nabla \times \nabla \times \boldsymbol{E}(\mathbf{r}) = i\omega^2 \, \underline{\underline{\mu}} \, \nabla \times \boldsymbol{H}(\mathbf{r}) \tag{7}$$

which yields the vector Helmholtz equation

$$\nabla \times \nabla \times \boldsymbol{E}(\mathbf{r}) = \omega^2 \underline{\underline{\mu}} \, \underline{\underline{\varepsilon}} \, \boldsymbol{E}(\mathbf{r}) \tag{8}$$

The left side of equation (8) can be expressed as

$$\nabla \times \nabla \times \boldsymbol{E}(\mathbf{r}) = \begin{vmatrix} \hat{\imath} & \hat{\jmath} & \hat{k} \\ \dfrac{\partial}{\partial x} & \dfrac{\partial}{\partial y} & \dfrac{\partial}{\partial z} \\ \dfrac{\partial}{\partial y} E_z - \dfrac{\partial}{\partial z} E_y & \dfrac{\partial}{\partial z} E_x - \dfrac{\partial}{\partial x} E_z & \dfrac{\partial}{\partial x} E_y - \dfrac{\partial}{\partial y} E_x \end{vmatrix} \tag{9}$$

Hence

$$\nabla \times \nabla \times \boldsymbol{E}(\mathbf{r}) = \left( \dfrac{\partial}{\partial y} \left\{ \dfrac{\partial}{\partial x} E_y - \dfrac{\partial}{\partial y} E_x \right\} - \dfrac{\partial}{\partial z} \left\{ \dfrac{\partial}{\partial z} E_x - \dfrac{\partial}{\partial x} E_z \right\} \right) \hat{\imath}$$
$$- \left( \dfrac{\partial}{\partial x} \left\{ \dfrac{\partial}{\partial x} E_y - \dfrac{\partial}{\partial y} E_x \right\} - \dfrac{\partial}{\partial z} \left\{ \dfrac{\partial}{\partial y} E_z - \dfrac{\partial}{\partial z} E_y \right\} \right) \hat{\jmath} \tag{10}$$
$$+ \left( \dfrac{\partial}{\partial x} \left\{ \dfrac{\partial}{\partial z} E_x - \dfrac{\partial}{\partial x} E_z \right\} - \dfrac{\partial}{\partial y} \left\{ \dfrac{\partial}{\partial y} E_z - \dfrac{\partial}{\partial z} E_y \right\} \right) \hat{k}$$

So we have

$$\nabla \times \nabla \times \boldsymbol{E}(\mathbf{r}) = \left( \left\{ \dfrac{\partial^2}{\partial yx} E_y - \dfrac{\partial^2}{\partial yy} E_x \right\} - \left\{ \dfrac{\partial^2}{\partial zz} E_x - \dfrac{\partial^2}{\partial zx} E_z \right\} \right) \hat{\imath}$$
$$- \left( \left\{ \dfrac{\partial^2}{\partial xx} E_y - \dfrac{\partial^2}{\partial xy} E_x \right\} - \left\{ \dfrac{\partial^2}{\partial zy} E_z - \dfrac{\partial^2}{\partial zz} E_y \right\} \right) \hat{\jmath} \tag{11}$$
$$+ \left( \left\{ \dfrac{\partial^2}{\partial xz} E_x - \dfrac{\partial^2}{\partial xx} E_z \right\} - \left\{ \dfrac{\partial^2}{\partial yy} E_z - \dfrac{\partial^2}{\partial yz} E_y \right\} \right) \hat{k}$$

Note that in the above equation, the y component is zero, therefore the left side of equation (8) will be

$$\nabla \times \nabla \times \boldsymbol{E}(\mathbf{r}) = \left( \dfrac{\partial^2}{\partial zx} E_z - \dfrac{\partial^2}{\partial zz} E_x \right) \hat{\imath} + \left( \left\{ \dfrac{\partial^2}{\partial xz} E_x - \dfrac{\partial^2}{\partial xx} E_z \right\} \right) \hat{k} \tag{12}$$

In the following we consider Gauss's law for neutral medium

$$\nabla \cdot \boldsymbol{D} = 0 \tag{13}$$

So equation (12) can be written as

$$\nabla \times \nabla \times \boldsymbol{E}(\mathbf{r}) = -\left(\frac{\partial^2}{\partial z^2} + \frac{\varepsilon_\parallel}{\varepsilon}\frac{\partial^2}{\partial x^2}\right)E_x\hat{\imath} - \left(\frac{\partial^2}{\partial x^2} + \frac{\varepsilon}{\varepsilon_\parallel}\frac{\partial^2}{\partial z^2}\right)E_z\hat{k} \qquad 14$$

The right side of equation (8) will be expressed as follows

$$\omega^2 \underline{\underline{\mu}}\,\underline{\underline{\varepsilon}}\,\boldsymbol{E}(\mathbf{r}) = \omega^2 \begin{bmatrix} \varepsilon_x & 0 & 0 \\ 0 & \varepsilon_\perp & 0 \\ 0 & 0 & \varepsilon_\perp \end{bmatrix}\begin{bmatrix} \mu_\perp & 0 & 0 \\ 0 & \mu_\parallel & 0 \\ 0 & 0 & \mu_\perp \end{bmatrix}[E_x \ E_y \ E_z] \qquad 15$$

Or

$$\omega^2 \underline{\underline{\mu}}\,\underline{\underline{\varepsilon}}\,\boldsymbol{E}(\mathbf{r}) = \omega^2 \varepsilon_x \mu_\perp E_x \hat{\imath} + \varepsilon_\perp \mu_\perp E_z \hat{k} \qquad 16$$

By substituting equations (14) and (16) in equation (8), the dispersion equation for the electric field is obtained [18, 19],

$$\varepsilon_\perp \beta^2 + \varepsilon_\parallel \alpha^2 = \omega^2 \varepsilon_\perp \varepsilon_\parallel \mu_\perp$$

Or

$$\frac{\beta^2}{\varepsilon_\parallel} + \frac{\alpha^2}{\varepsilon_\perp} = \omega^2 \mu_\perp \qquad 17$$

Analogously, for magnetic modes we have

$$\mu_\parallel \alpha^2 + \mu_\perp \beta^2 = \omega^2 \mu_\parallel \varepsilon_\perp \mu_\perp$$

Or

$$\frac{\beta^2}{\mu_\parallel} + \frac{\alpha^2}{\mu_\perp} = \omega^2 \varepsilon_\perp \qquad 18$$

Now we are able to analyze the dispersion equation in dielectric-magnetic mediums by using relations (17) and (18).

3. **SIMULATION AND ANALYSIS**

In this section all calculations are done by MATLAB software. We investigate the dispersion of electromagnetic waves (by changing the signs of elements of the permittivity and permeability tensors) in dissipative and non-dissipative uniaxial dielectric-magnetic mediums by using equations (17) and (18). In the following figures, the black, red, green and blue curves are related to the real part of β for non-dissipative, the imaginary part of β for non-dissipative, the real part of β for dissipative, and the real imaginary part of β for dissipative mediums, respectively. Comparing the dissipative and non-dissipative cases, the dissipation effects are clearly seen. It can be seen that in dissipation, the allowed cases are reduced to eight cases. There are situations that no allowed relation for the non-dissipative case exists, but there is an answer for the same situation in the case of dissipation. All figures are for electric polarization. In the following, we have shown other cases for electric and magnetic modes in table 1 and table 2, respectively [20, 21].

First, we consider the case with $\varepsilon_\perp > 0$, $\mu_\perp > 0$, $\varepsilon_\parallel > 0$ (figure 1) for non-dissipative medium and $Re\{\varepsilon_\perp\} > 0$, $Re\{\mu_\perp\} > 0$, $Re\{\varepsilon_\parallel\} > 0$ for dissipative medium for specific values of $\varepsilon$ and $\mu$ [18]. In figure 1 dispersion surface is ellipsoid for real part of β for non-dissipative medium (black curve) and hyperboloid of two sheets for imaginary part of β for non-dissipative medium (red curve).

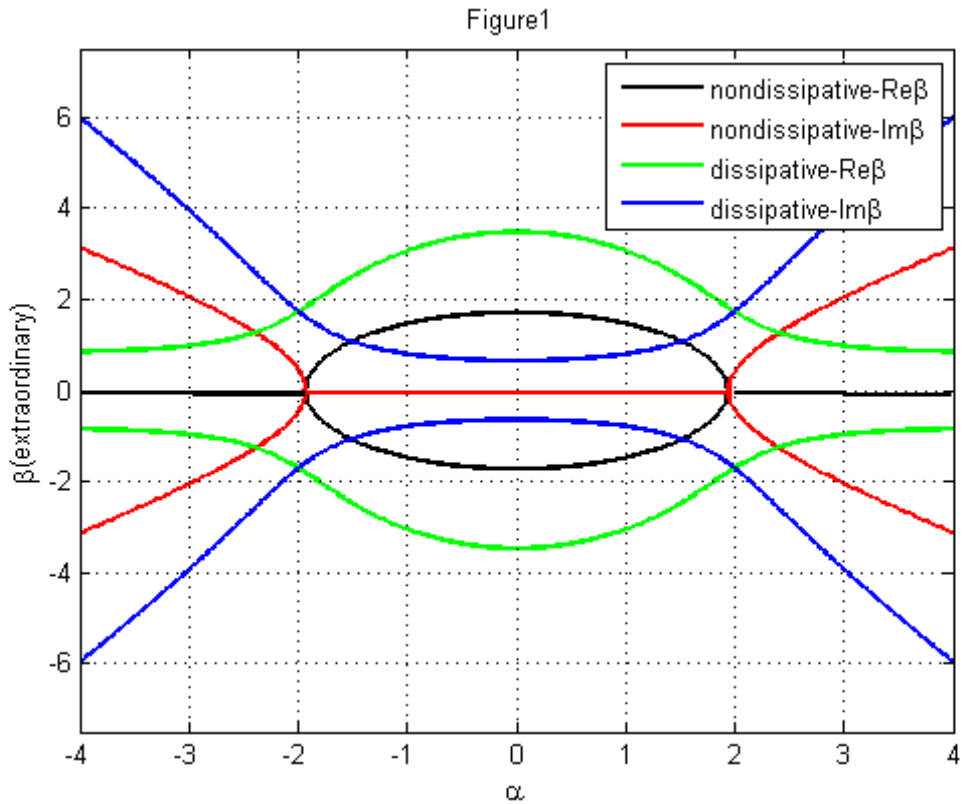

Figure 1: First case $\varepsilon_\perp$=2.5$\varepsilon_0$, $\varepsilon_\parallel$=2$\varepsilon_0$, $\mu_\perp$=1.5$\mu_0$ (non-dissipative medium), Plot for the real part (black curve) and imaginary part (red curve) of β. $\varepsilon_\perp$= (2+5i) $\varepsilon_0$, $\varepsilon_\parallel$= (6+0.75i) $\varepsilon_0$, $\mu_\perp$= (2+0.5i) $\mu_0$ (dissipative medium), Plot for the real part (green curve) and imaginary part (blue curve) of β. The values of $\alpha$ and β are normalized.

Secondly, we consider the case with $\varepsilon_\perp < 0$, $\mu_\perp > 0$, $\varepsilon_\parallel < 0$ (figure 2) for non-dissipative medium and $Re\{\varepsilon_\perp\} < 0$, $Re\{\mu_\perp\} > 0$, $Re\{\varepsilon_\parallel\} < 0$ for dissipative medium for specific values of $\varepsilon$ and $\mu$. In figure 2 we have dispersion surface only for imaginary part of β for non-dissipative medium (red curve). The above conditions are not allowed to dispersion for real part of β for non-dissipative medium, also we can obviously see that there is no answer for the real and imaginary part of β for dissipative media.

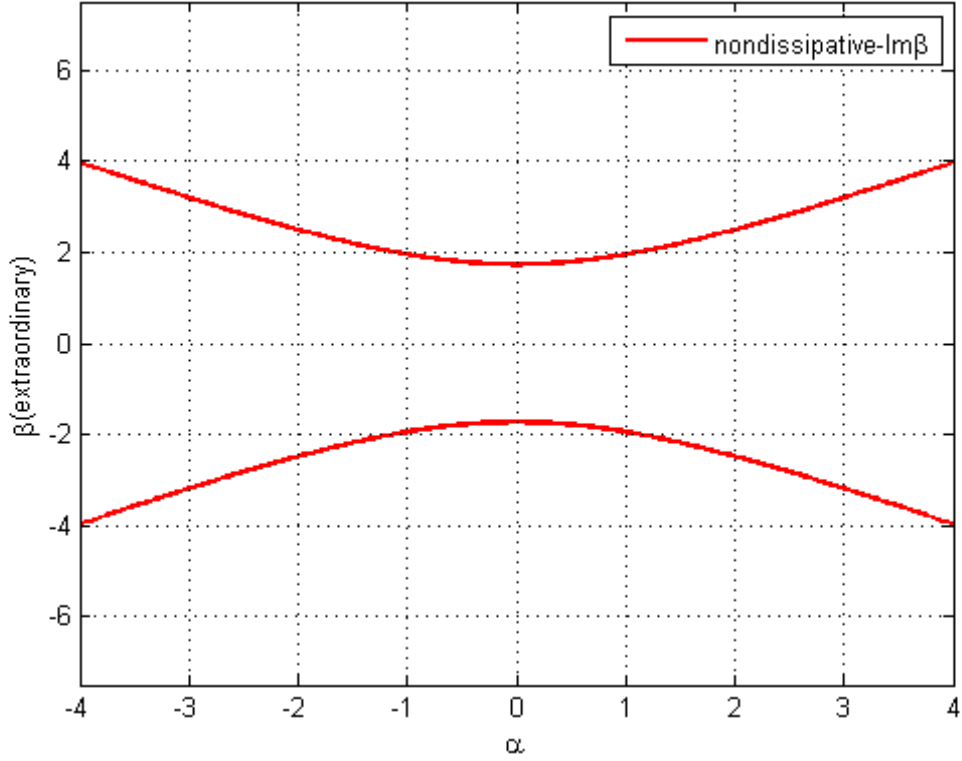

Figure 2: Second case $\varepsilon_\perp$=-2.5$\varepsilon_0$, $\varepsilon_\parallel$=-2$\varepsilon_0$, $\mu_\perp$=1.5$\mu_0$ (non-dissipative medium), Plot for the imaginary part of β. The real part of β is not allowed in a non-dissipative medium. $\varepsilon_\perp$= (-2+5i) $\varepsilon_0$, $\varepsilon_\parallel$= (-6+0.75i) $\varepsilon_0$, $\mu_\perp$= (2+0.5i) $\mu_0$ (dissipative medium), both real and imaginary part of β are not allowed. The values of $\alpha$ and β are normalized.

Thirdly, we consider the case with $\varepsilon_\perp > 0$, $\mu_\perp < 0$, $\varepsilon_\parallel > 0$ for non-dissipative medium and $Re\{\varepsilon_\perp\} > 0$, $Re\{\mu_\perp\} < 0$, $Re\{\varepsilon_\parallel\} > 0$ for dissipative medium for specific values of $\varepsilon$ and $\mu$. In figure 3 we can obviously see that there is no solution for real part of β for non-dissipative medium (i.e. nonpropagating). For dissipative modes, there are values for both real and imaginary part of β.

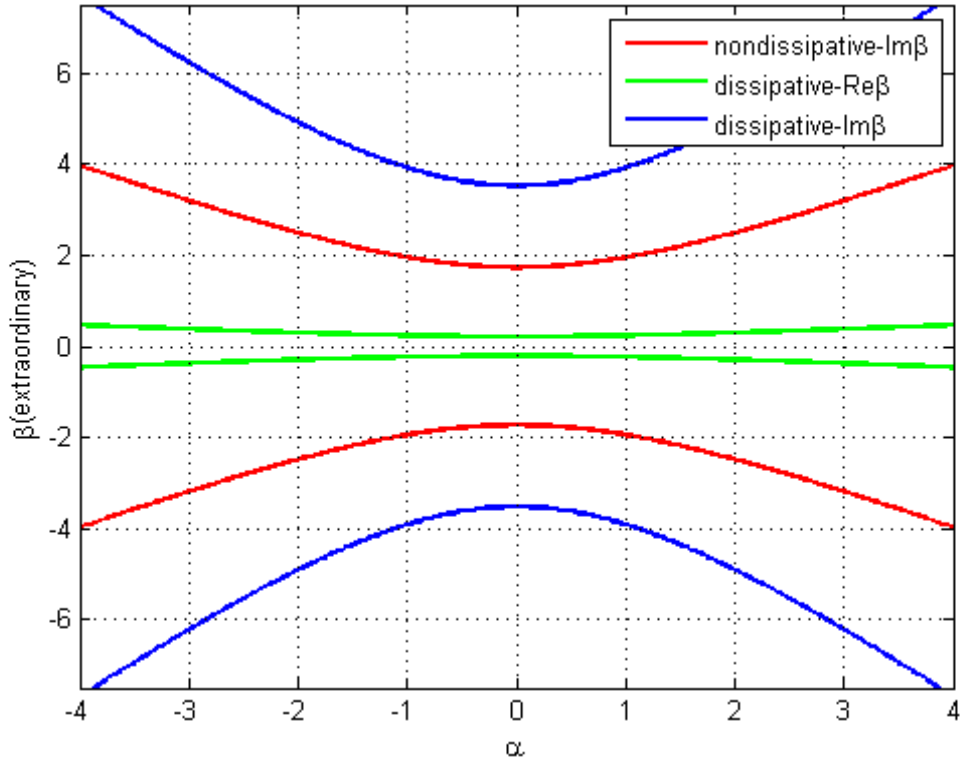

**Figure 3:** Third case $\varepsilon_\perp=2.5\varepsilon_0$, $\varepsilon_\parallel=2\varepsilon_0$, $\mu_\perp=-1.5\mu_0$ (non-dissipative medium), Plot for the imaginary part of β (red curve). The real part of β is not allowed in a non-dissipative medium. $\varepsilon_\perp$= (2+5i) $\varepsilon_0$, $\varepsilon_\parallel$= (6+0.75i) $\varepsilon_0$, $\mu_\perp$= (-2+0.5i) $\mu_0$ (dissipative medium), Plot for the real part (green curve) and imaginary part (blue curve) of β. The values of $\alpha$ and β are normalized.

Fourthly, we consider the case with $\varepsilon_\perp < 0$, $\mu_\perp < 0$, $\varepsilon_\parallel < 0$ (figure 4) for non-dissipative medium and $Re\{\varepsilon_\perp\} < 0$, $Re\{\mu_\perp\} < 0$, $Re\{\varepsilon_\parallel\} < 0$ for dissipative medium for specific values of $\varepsilon$ and $\mu$. In figure 4 dispersion surface is ellipsoid for real part of β for non-dissipative medium (black curve) and dispersion surface is hyperboloid of two sheets for imaginary part of β for non-dissipative medium (red curve). We can see that there is no solution for dissipative medium (i.e. nonpropagating).

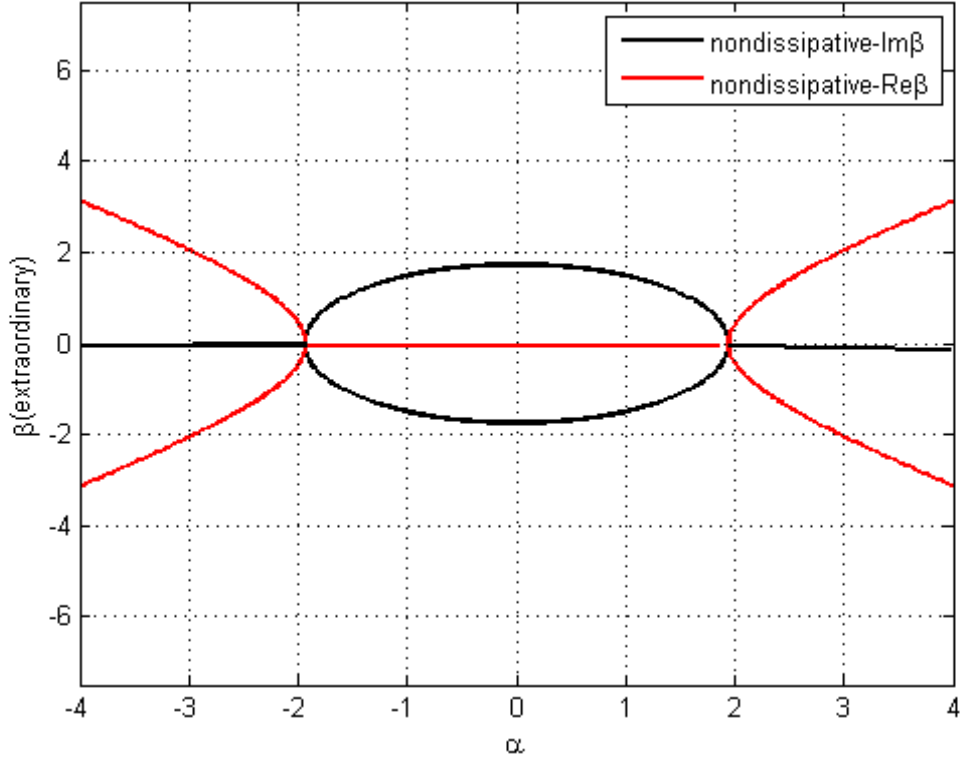

Figure 4: Fourth case $\varepsilon_\perp=-2.5\varepsilon_0$, $\varepsilon_\parallel=-2\varepsilon_0$, $\mu_\perp=-1.5\mu_0$ (for non-dissipative), (non-dissipative medium), Plot for the real part (black curve) and imaginary part (red curve) of β. $\varepsilon_\perp=$ (-2+5i) $\varepsilon_0$, $\varepsilon_\parallel=$ (-6+0.75i) $\varepsilon_0$, $\mu_\perp=$ (2+0.5i) $\mu_0$ (dissipative medium), both real and imaginary part of β are not allowed in dissipative mediums. The values of $\alpha$ and β are normalized

We have shown other cases for electric and magnetic modes in table 1 and table 2, respectively. $\varepsilon_\parallel$, $\varepsilon_\perp$, and $\mu_\perp, \mu_\parallel$, which are the elements of the permittivity and permeability tensors for non-dissipative modes and $\text{Re}\{\varepsilon_\parallel\}$, $Re\{\varepsilon_\perp\}$, and $Re\{\mu_\perp\}, Re\{\mu_\parallel\}$, which are the elements of the permittivity and permeability tensors for real part of β for dissipative modes. Note that we plotted just the first four cases for electric modes.

**Table 1: Different modes of dispersion for non-dissipative and dissipative medium (Electric polarization).**

| | | Electric modes | | | |
|---|---|---|---|---|---|
| | | non-dissipative | | dissipative | |
| | | Re {β} | Im {β} | Re {β} | Im {β} |
| 1 | $\varepsilon_\perp$ and $Re\varepsilon_\perp > 0$<br>$\varepsilon_\parallel$ and $Re\varepsilon_\parallel > 0$<br>$\mu_\perp$ and $Re\mu_\perp > 0$ | Ellipsoid<br><br>Figure 1-black curve | Hyperboloid of two sheets<br>Figure 1- red curve | Propagating<br><br>Figure 1- green curve | Propagating<br><br>Figure 1-blue curve |
| 2 | $\varepsilon_\perp$ and $Re\varepsilon_\perp < 0$<br>$\varepsilon_\parallel$ and $Re\varepsilon_\parallel < 0$<br>$\mu_\perp$ and $Re\mu_\perp > 0$ | Nonpropagating | Hyperboloid of one sheets<br>Figure 2- red curve | Nonpropagating | Nonpropagating |
| 3 | $\varepsilon_\perp$ and $Re\varepsilon_\perp > 0$<br>$\varepsilon_\parallel$ and $Re\varepsilon_\parallel > 0$<br>$\mu_\perp$ and $Re\mu_\perp < 0$ | Nonpropagating | Hyperboloid of one sheets<br>Figure 3- red curve | Propagating<br><br>Figure 3- green curve | Propagating<br><br>Figure 1-blue curve |
| 4 | $\varepsilon_\perp$ and $Re\varepsilon_\perp < 0$<br>$\varepsilon_\parallel$ and $Re\varepsilon_\parallel < 0$<br>$\mu_\perp$ and $Re\mu_\perp < 0$ | Ellipsoid<br><br>Figure 4-black curve | Hyperboloid of two sheets<br>Figure 4- red curve | Nonpropagating | Nonpropagating |
| 5 | $\varepsilon_\perp$ and $Re\varepsilon_\perp < 0$<br>$\varepsilon_\parallel$ and $Re\varepsilon_\parallel > 0$<br>$\mu_\perp$ and $Re\mu_\perp > 0$ | Hyperboloid of one sheets | Nonpropagating | propagating | propagating |
| 6 | $\varepsilon_\perp$ and $Re\varepsilon_\perp > 0$<br>$\varepsilon_\parallel$ and $Re\varepsilon_\parallel < 0$<br>$\mu_\perp$ and $Re\mu_\perp > 0$ | Hyperboloid of two sheets | Ellipsoid | Nonpropagating | Nonpropagating |
| 7 | $\varepsilon_\perp$ and $Re\varepsilon_\perp < 0$<br>$\varepsilon_\parallel$ and $Re\varepsilon_\parallel > 0$<br>$\mu_\perp$ and $Re\mu_\perp < 0$ | Hyperboloid of two sheets | Ellipsoid | propagating | propagating |
| 8 | $\varepsilon_\perp$ and $Re\varepsilon_\perp > 0$<br>$\varepsilon_\parallel$ and $Re\varepsilon_\parallel < 0$<br>$\mu_\perp$ and $Re\mu_\perp < 0$ | Hyperboloid of one sheets | Nonpropagating | Nonpropagating | Nonpropagating |

**Table 2: Different modes of dispersion for non-dissipative and dissipative medium (magnetic polarization).**

| | | magnetic modes | | | |
|---|---|---|---|---|---|
| | | non-dissipative | | dissipative | |
| | | Re {β} | Im {β} | Re {β} | Im {β} |
| 1 | $\mu_\perp$ and $Re\,\mu_\perp > 0$<br>$\mu_\parallel$ and $Re\mu_\parallel > 0$<br>$\varepsilon_\perp$ and $Re\varepsilon_\perp > 0$ | Ellipsoid | Hyperboloid of two sheets | Propagating | Propagating |
| 2 | $\mu_\perp$ and $Re\,\mu_\perp < 0$<br>$\mu_\parallel$ and $Re\mu_\parallel < 0$<br>$\varepsilon_\perp$ and $Re\varepsilon_\perp > 0$ | Nonpropagating | Hyperboloid of one sheets | Nonpropagating | Nonpropagating |
| 3 | $\mu_\perp$ and $Re\,\mu_\perp > 0$<br>$\mu_\parallel$ and $Re\mu_\parallel > 0$<br>$\varepsilon_\perp$ and $Re\varepsilon_\perp < 0$ | Nonpropagating | Hyperboloid of one sheets | Propagating | Propagating |
| 4 | $\mu_\perp$ and $Re\,\mu_\perp < 0$<br>$\mu_\parallel$ and $Re\mu_\parallel < 0$<br>$\varepsilon_\perp$ and $Re\varepsilon_\perp < 0$ | Ellipsoid | Hyperboloid of two sheets | Nonpropagating | Nonpropagating |
| 5 | $\mu_\perp$ and $Re\,\mu_\perp < 0$<br>$\mu_\parallel$ and $Re\mu_\parallel > 0$<br>$\varepsilon_\perp$ and $Re\varepsilon_\perp > 0$ | Hyperboloid of one sheets | Nonpropagating | propagating | propagating |
| 6 | $\mu_\perp$ and $Re\,\mu_\perp > 0$<br>$\mu_\parallel$ and $Re\mu_\parallel < 0$<br>$\varepsilon_\perp$ and $Re\varepsilon_\perp > 0$ | Hyperboloid of two sheets | Ellipsoid | Nonpropagating | Nonpropagating |
| 7 | $\mu_\perp$ and $Re\,\mu_\perp < 0$<br>$\mu_\parallel$ and $Re\mu_\parallel > 0$<br>$\varepsilon_\perp$ and $Re\varepsilon_\perp < 0$ | Hyperboloid of two sheets | Ellipsoid | propagating | propagating |
| 8 | $\mu_\perp$ and $Re\,\mu_\perp > 0$<br>$\mu_\parallel$ and $Re\mu_\parallel < 0$<br>$\varepsilon_\perp$ and $Re\varepsilon_\perp < 0$ | Hyperboloid of one sheets | Nonpropagating | Nonpropagating | Nonpropagating |

## 4. CONCLUSION

In this work, considering phenomenon of dissipation, propagation of electromagnetic planar waves in negative refractive index uniaxial mediums has been studied. We showed that dispersion relation in a non-dissipative medium is elliptical, hyperboloid of one sheet and hyperboloid of two sheets; but in the case of a dissipative medium this order vanishes. In general, according to the signs of elements of dielectric permittivity and magnetic permeability tensors, sixteen possible states are obtained for propagation in a non-dissipative electromagnetic medium. In dissipation, the number of states is reduced and reaches eight states. There are conditions in which there is no possible relation for propagation in a non-dissipative medium for real or imaginary parts of β, but there is an answer for the same condition in a dissipative medium.